\documentclass[english]{article}
\usepackage[T1]{fontenc}
\usepackage[latin9]{inputenc}
\usepackage[paperwidth=20cm]{geometry}
\usepackage{amsthm}
\usepackage{amsmath}
\usepackage{amssymb}
\usepackage[all]{xy}

\makeatletter
\newcommand{\lyxaddress}[1]{
\par {\raggedright #1
\vspace{1.4em}
\noindent\par}
}
\theoremstyle{plain}
\newtheorem{thm}{\protect\theoremname}
  \theoremstyle{definition}
  \newtheorem{defn}[thm]{\protect\definitionname}
  \theoremstyle{definition}
  \newtheorem*{example*}{\protect\examplename}
  \theoremstyle{plain}
  \newtheorem{prop}[thm]{\protect\propositionname}

\makeatother

\usepackage{babel}
  \providecommand{\definitionname}{Definition}
  \providecommand{\examplename}{Example}
  \providecommand{\propositionname}{Proposition}
\providecommand{\theoremname}{Theorem}

\begin{document}

\title{Defining relations and flip Dynkin Superdiagram}

\author{B. Ransingh}
\date{}

\maketitle

\lyxaddress{\begin{center}
Department of Mathematics \\National Institue of Technology\\ Rourkela,
India-769008\\email- bransingh@gmail.com
\par\end{center}}

AMS 2010 Subject classification Scheme- 17B60

Keyword- Lie superalgebras, Dynkin diagrams
\begin{abstract}
The motivation comes from boson fermion correspondence. This article
shows for each fermionic root there is correspondence a bosonic root,
as a result we get for each Dynkin diagram of Lie Superalgebra a corresoponding
flip Dynkin Superdiagram. This article construct all the filp Dynkin
Superdiagrams of Lie superalgebras(LS). This can create non conjugate
classes Borel subalgebra (subsuperalgebras) or non isomorphic Dynkin
diagrams of LS using $\in\delta$ sequences. We have got the defining
relations for the both the Dynkin diagrams and flip Dynkin diagrams.
\end{abstract}

\section{Introduction and Preliminaries}

The boson-fermion correspondence is an isomorphism between two representations
of an infinite-dimensional Heisenberg algebra (or the closely related
oscillator algebra). These representations are on the bosonic and
fermionic Fock spaces.  The supersymmetry
for this process was introduced in particle physics, its physical
consequences follows the algebraical structure, so called graded
Lie algebras or Lie superalgebra. {[}5{]}
\begin{defn}
A Lie superalgebras $g$ is an algebra graded over $\mathbb{Z}_{2}$
, i.e., $g$ is a direct sum of vector spaces $g=g_{0}\oplus g_{1}$,
and such that the bracket satisfies \end{defn}
\begin{enumerate}
\item $[g_{i},g_{j}]\subset g_{i+j(mod2)}$
\item $[x_{i},y_{i}]=-(-1)^{ij}[y_{j},x_{i}]$, $\forall$ $x_{i}\in g_{i}$
, $y_{j}\in g_{j}$; (Skew Supersymmetry)
\item $[x_{i},[y_{j},z]]=[[x_{i},y_{j}],z]+\left(-1\right)^{ij}[y_{j},[x_{i},z]]$, $\forall z\in g$
(Super Jacobi Identity)
\end{enumerate}
The circles in the Dynkin diagram of Lie superalgebras $\bigcirc,$
$\otimes$ and $\bullet$ are respectively called white, grey and
black. $\otimes$ is also denoted as an isotropic odd simple root.
Each simple bosonic root is associated with a white circle $\bigcirc$, 
each fermionic root is associated with a grey root $\otimes$ if $a_{ij}=0$
and $a_{ij}\neq0$ a black dot $\bullet$.

For general linear Lie superalgebra  $gl(m|n)$ root system defined
by $\triangle$, set $\triangle^{+}$ of positive root , $\triangle^{-}$
negative roots and a set $\Pi$ of simple roots in $\triangle^{+}$.
A root $\alpha$ is even if $g_{\alpha}\subset g_{\overline{0}}$
and it is odd if $g_{\alpha}\subset g_{\overline{1}}$. Thus $\triangle_{\overline{0}}$
and $\triangle_{\overline{1}}$ are the set of even and odd root respectivily.{[}1{]}{[}2{]}{[}3{]}{[}4{]}

\section{Flip Dynkin Super diagrams and Flip Cartan Matrix}
\begin{defn}
Fliping the fermions and bosonic root gives a filp Dynkin superdiagram.
Flip is a map $f$

\[
f:\begin{array}{c}
\bigcirc\rightarrow\otimes\\
\otimes\rightarrow\bigcirc
\end{array}
\]

If there is no more than one line between the roots of the Dynkin
diagrams (simply laced Dynkin diagram). Thus fliping preserve the
line and orientation of doubly and triply connected Dynkin diagram.
Flip Dynkin Superdiagram creats flip Cartan matrix.
\end{defn}
Lie Superalgebra $sl(m,n)$ case

\[
\underset{e_{1}-e_{2}}{\bigcirc}-\underset{e_{2}-e_{3}}{\bigcirc}-\cdots-\underset{e_{m}-e_{m+1}}{\bigcirc}
-\underset{e_{m+1}-\delta_{1}}{\otimes}-\underset{\delta_{1}-\delta_{2}}{\bigcirc}-\cdots-
\underset{\delta_{n-1}-\delta_{n}}{\bigcirc}-\underset{\delta_{n-1}-\delta_{n}}{\bigcirc}
\]

the corresponding cartan matrix
\[
\left(\begin{array}{ccccccc}
2 & -1\\
-1 & \ddots & \ddots\\
 & \ddots & 2 & -1\\
 &  & -1 & 0 & 1\\
 &  &  & 1 & -2 & \ddots\\
 &  &  &  & \ddots & \ddots & 1\\
 &  &  &  &  & 1 & -2
\end{array}\right)
\]

$\xymatrix{ & \underset{e_{1}-e_{2}}{\bigcirc}-\underset{e_{2}-e_{3}}
{\bigcirc}-\cdots-\underset{e_{m}-e_{m+1}}{\bigcirc}-\underset{e_{m+1}
-\delta_{1}}{\otimes}-\underset{\delta_{1}-\delta_{2}}{\bigcirc}-\cdots
-\underset{\delta_{n-1}-\delta_{n}}{\bigcirc}-\underset{\delta_{n-1}-\delta_{n}}{\bigcirc}\ar[d]_{f}\\
 & \otimes-\otimes-\cdots-\otimes-\bigcirc-\otimes-\cdots-\otimes-\otimes}$

So the supersymmetry partner of $sl(m,n)$ is the flip Dynkin Superdiagram

\[
\otimes-\otimes-\cdots-\otimes-\bigcirc-\otimes-\cdots-\otimes-\otimes
\]

correspond cartan matrix with the normalisation 
$(e_{i},e_{j})=\delta_{ij}(i,j=1,\cdots,m+1)$, $(\delta_{k},\delta_{l})=-\delta_{kl}(k,l=1,\cdots,n+1)$,
$(e_{i},\delta_{k})=0$ and the flip Cartan matrix become

\[
\left(\begin{array}{ccccccc}
2 & -1\\
-1 & \ddots & \ddots\\
 & \ddots & 2 & -1\\
 &  & -1 & 0 & 1\\
 &  &  & 1 & -2 & \ddots\\
 &  &  &  & \ddots & \ddots & 1\\
 &  &  &  &  & 1 & -2
\end{array}\right)\begin{array}{c}
\\
\underrightarrow{f}
\end{array}\left(\begin{array}{ccccccc}
0 & -1\\
-1 & \ddots & \ddots\\
 & \ddots & 0 & -1\\
 &  & -1 & -2 & 1\\
 &  &  & 1 & 0 & \ddots\\
 &  &  &  & \ddots & \ddots & -1\\
 &  &  &  &  & -1 & 0
\end{array}\right)
\]

\begin{example*}
$sl(2,2)$

The Dynkin diagram
\end{example*}
\[
\underset{e_{1}-e_{2}}{\bigcirc}-\underset{e_{2}-e_{3}}{\bigcirc}-\underset{e_{3}-\delta_{1}}{\otimes}-
\underset{\delta_{1}-\delta_{2}}{\bigcirc}-\underset{\delta_{2}-\delta_{3}}{\bigcirc}
\]

Corresponding Cartan matrix
\[
\left(\begin{array}{ccccc}
2 & -1 & 0 & 0 & 0\\
-1 & 2 & -1 & 0 & 0\\
0 & -1 & 0 & 1 & 0\\
0 & 0 & -1 & -2 & 1\\
0 & 0 & 0 & 1 & -2
\end{array}\right)
\]

The associated flip dynkin superdiagram

\[
\underset{e_{1}-\delta_{1}}{\otimes}-\underset{\delta_{1}-e_{2}}{\otimes}-\underset{e_{2}-e_{3}}
{\bigcirc}-\underset{e_{3}-\delta_{2}}{\otimes}-\underset{\delta_{2}-e_{4}}{\otimes}
\]

Corresponding Cartan Matrix

\[
CM1=\left(\begin{array}{ccccc}
2 & -1 & 0 & 0 & 0\\
-1 & 2 & -1 & 0 & 0\\
0 & -1 & 0 & 1 & 0\\
0 & 0 & -1 & -2 & 1\\
0 & 0 & 0 & 1 & -2
\end{array}\right)\begin{array}{c}
\\
\underrightarrow{f} = CM2 =
\end{array}\left(\begin{array}{ccccc}
0 & -1 & 0 & 0 & 0\\
-1 & 0 & 1 & 0 & 0\\
0 & -1 & 2 & -1 & 0\\
0 & 0 & -1 & 0 & 1\\
0 & 0 & 0 & 1 & 0
\end{array}\right)
\]

\subsection{Serre-type relations. }

$[H_{i},H_{i}]=0$, $[X_{i}^{+},X_{j}^{-}]=\delta_{ij}H_{i}$,$[H_{i},X_{j}^{\pm}]=\pm a_{ij}X_{j}^{\pm}$
for $H_{i}=[X_{i}^{+},X_{i}^{-}]$
\begin{prop}\cite{grozman1} The number $k_{in}$ and $k_{ni}$ are expressed
in terms of $B_{kj}$ as follows

\[
\mbox{(ad}_{X_{k}^{\pm}}\mbox{)}^{1+B_{kj}}(X_{j}^{\pm})=0\quad\mbox{for }k\neq j
\]

\[
[X_{i}^{\pm},X_{i}^{\pm}]=0\quad\mbox{if }A_{ii}=0
\]

The above relations will be called Serre relations of the Lie superalgebras.

Here 
\[
B_{kj}=\begin{cases}
\begin{array}{cc}
-\frac{2A_{kj}}{A_{kk}} & \mbox{if }A_{kk}\neq0\mbox{ and }-\frac{2A_{kj}}{A_{kk}}\in\mathbb{Z}_{+},\\
1 & \mbox{if }i_{k}=\bar{1},A_{kk}=0,A_{kj}\neq0\\
0 & \mbox{if }i_{k}=\bar{1},A_{kk}=A_{kj}=0\\
0 & \mbox{if }i_{k}=\bar{0},A_{kk}=0,A_{kj}=0
\end{array}\end{cases}
\]

\end{prop}
Defining relations for Cartan matrix CM1 

$[x_{1},x_{3}]=0$

$[x_{1},x_{4}]=0$

$[x_{1},x_{5}]=0$

$[x_{2},x_{4}]=0$

$[x_{3},x_{3}]=0$

$[x_{3},x_{5}]=0$

$[x_{1},[x_{1},x_{2}]]=0$

$[x_{2},[x_{1},x_{2}]]=0$

$[x_{2},[x_{2},x_{3}]]=0$

$[x_{4},[x_{4},x_{5}]]=0$

$[[x_{2},x_{3}],[x_{3},x_{4}]]=0$

$[\{x_{1},x_{2},x_{3},x_{4},x_{5},[x_{1},x_{2}],[x_{2},x_{3}],[x_{3},x_{4}],[x_{4},x_{5}],[x_{3}[x_{1},x_{2}]]$

$[x_{4}[x_{2},x_{3}]]$,$[x_{4}[x_{3},x_{4}]]$,$[x_{5}[x_{3},x_{4}]]$,$[x_{4}[x_{4},[x_{2},x_{3}]]]$,$[x_{4}[x_{4},[x_{3},x_{4}]]]$,
$[[x_{1},x_{2}],[x_{3},x_{4}]]$,

$[[x_{2},x_{3}],[x_{3},x_{4}]]$,$[[x_{2},x_{3}],[x_{4},x_{5}]]$,$[[x_{3},x_{4}],[x_{3},x_{4}]]$,$[[x_{3},x_{4}],[x_{4},x_{5}]]$,

$[x_{4},[x_{4},[x_{4},[x_{2},x_{3}]]]]$,$[x_{4},[x_{4},[x_{4},[x_{3},x_{4}]]]]$,$[x_{1},x_{2}],[x_{4},[x_{4},[x_{3},x_{4}]]]]$,

$[x_{3},x_{4}],[x_{4},[x_{4},[x_{2},x_{3}]]]]$,$[x_{3},x_{4}],[x_{4},[x_{4},[x_{3},x_{4}]]]]$,

$[[x_{4},x_{5}],[x_{4},[x_{4},[x_{2},x_{3}]]]]$,$[[x_{4},x_{5}],[x_{4},[x_{4},[x_{2},x_{3}]]]]$,$[[x_{4},x_{5}],[x_{4},[x_{4},[x_{3},x_{4}]]]]$,

$[[x_{4},x_{5}],[x_{1},x_{2}],[x_{3},x_{4}]]]]$,

$[[x_{4},x_{5}],[x_{3},x_{4}],[x_{4},x_{5}]]]]$,$[[x_{4},x_{5}],[x_{3},x_{4}],[x_{4},x_{5}]]]]$,$[[x_{4},x_{5}],[x_{3},x_{4}],[x_{3},x_{4}]]]]$,

$[[x_{4},x_{5}],[x_{3},x_{4}],[x_{3},x_{4}]]]]$,

$[[x_{4},x_{5}],[x_{3},x_{4}],[x_{4},x_{5}]]]]$,

$[[x_{3},[x_{1},x_{2}]],[x_{4},[x_{3},x_{4}]]]$,$[[x_{3},[x_{1},x_{2}]],[x_{4},[x_{3},x_{4}]]]$,$[[x_{4},[x_{2},x_{3}]],$

$[x_{4},[x_{2},x_{3}]]]$,$[[x_{4},[x_{2},x_{3}]],[x_{4},[x_{3},x_{4}]]]$,

$[[x_{4},[x_{2},x_{3}]],[x_{5},[x_{3},x_{4}]]]$,$[[x_{4},[x_{3},x_{4}]],[x_{4},[x_{3},x_{4}]]]$,$[[x_{4},[x_{3},x_{4}]],$

$[x_{5},[x_{3},x_{4}]]]$,$[[x_{5},[x_{3},x_{4}]],[x_{5},[x_{3},x_{4}]]]$

Defining relations of Cartan Matrix CM2

$[x_{1},x_{1}]=0$

$[x_{1},x_{3}]=0$

$[x_{1},x_{4}]=0$

$[x_{1},x_{5}]=0$

$[x_{2},x_{2}]=0$

$[x_{2},x_{4}]=0$

$[x_{2},x_{5}]=0$

$[x_{3},x_{5}]=0$

$[x_{4},x_{4}]=0$

$[x_{5},x_{5}]=0$

$[x_{3},[x_{2},x_{3}]]=0$

$[x_{3},[x_{3},x_{4}]]=0$

$[x_{3},[x_{2},x_{3}]]=0$

$[[x_{3},x_{4}],[x_{4},x_{5}]]=0$

$[\{x_{1},x_{2},x_{3},x_{4},x_{5},[x_{1},x_{2}],[x_{2},x_{3}],[x_{3},x_{4}],[x_{4},x_{5}],[x_{3},[x_{1},x_{2}]]$,

$[x_{4},[x_{2},x_{3}]]$,$[x_{5},[x_{3},x_{4}]]$,$[[x_{1},x_{2}],[x_{2},x_{3}]]$,$[[x_{1},x_{2}],[x_{3},x_{4}]]$,

$[[x_{2},x_{3}],[x_{4},x_{5}]]$,$[[x_{1},x_{2}],[x_{3},[x_{1},x_{2}]]]$,$[[x_{1},x_{2}],[x_{4},[x_{2},x_{3}]]]$,

$[[x_{4},x_{5}],[x_{3},[x_{1},x_{2}]]]$,$[[x_{1},x_{2}],[[x_{1},x_{2}],[x_{2},x_{3}]]]$,$[[x_{1},x_{2}],$

$[[x_{1},x_{2}],[x_{3},x_{4}]]]$,$[[x_{4},x_{5}],[[x_{1},x_{2}],[x_{2},x_{3}]]]$,

$[[x_{3},[x_{1},x_{2}]],[x_{4},[x_{2},x_{3}]]]\}]$

For $sl(2,1)$ dynkin diagram$\underset{e_{1}-e_{2}}{\bigcirc}-\underset{e_{2}-e_{3}}{\bigcirc}-\underset{e_{3}-\delta_{1}}{\otimes}-\underset{\delta_{1}-\delta_{2}}{\bigcirc}$,
the filp dynkin superdiagram becomes $\underset{e_{1}-\delta_{1}}{\otimes}-\underset{\delta_{1}-e_{2}}{\otimes}-\underset{e_{2}-\delta_{2}}{\bigcirc}-\underset{\delta_{2}-e_{3}}{\otimes}$
. 

For $sl(3,2)$ dynkin diagram$\underset{e_{1}-e_{2}}{\bigcirc}-\underset{e_{2}-e_{3}}{\bigcirc}-\underset{e_{3}-e_{4}}{\bigcirc}-\underset{e_{4}-\delta_{1}}{\otimes}-\underset{\delta_{1}-\delta_{2}}{\bigcirc}$,
the filp dynkin superdiagram becomes $\underset{e_{1}-\delta_{1}}{\otimes}-\underset{\delta_{1}-e_{2}}{\otimes}-\underset{e_{2}-\delta_{2}}{\bigcirc}-\underset{\delta_{2}-e_{3}}{\otimes}$
.

For $sl(5,5)$ dynkin diagram 
\[
\underset{e_{1}-e_{2}}{\bigcirc}-\underset{e_{2}-e_{3}}{\bigcirc}-\underset{e_{3}-e_{4}}{\bigcirc}-\underset{e_{4}-e_{5}}{\bigcirc}-\underset{e_{5}-e_{6}}{\bigcirc}-\underset{e_{6}-\delta_{1}}{\otimes}-\underset{\delta_{1}-\delta_{2}}{\bigcirc}-\underset{\delta_{2}-\delta_{3}}{\bigcirc}-\underset{\delta_{3}-\delta_{4}}{\bigcirc}-\underset{\delta_{4}-\delta_{5}}{\bigcirc}-\underset{\delta_{5}-\delta_{6}}{\bigcirc}
\]
corresponding flip diagram 
\[
\underset{e_{1}-\delta_{1}}{\otimes}-\underset{\delta_{1}-e_{2}}{\otimes}-\underset{e_{2}-\delta_{2}}{\otimes}-\underset{\delta_{2}-e_{3}}{\otimes}-\underset{e_{3}-\delta_{3}}{\otimes}-\underset{\delta_{3}-\delta_{4}}{\bigcirc}-\underset{\delta_{4}-e_{4}}{\otimes}-\underset{e_{4}-\delta_{5}}{\otimes}-\underset{\delta_{5}-e_{5}}{\otimes}-\underset{e_{5}-\delta_{6}}{\otimes}-\underset{\delta_{6}-e_{6}}{\otimes}
\]
 $osp(m,2n)$ case $\bigcirc-\bigcirc-\cdots\bigcirc-\otimes-\bigcirc-\cdots-\bigcirc\Rightarrow\bigcirc$

\[
\underset{e_{1}-e_{2}}{\bigcirc}-\underset{e_{2}-e_{3}}{\bigcirc}-\cdots-\underset{e_{m}-e_{m+1}}{\bigcirc}-\underset{e_{m+1}-\delta_{1}}{\otimes}-\underset{\delta_{1}-\delta_{2}}{\bigcirc}-\cdots-\underset{\delta_{n-1}-\delta_{n}}{\bigcirc}\Rightarrow\underset{\delta_{n-1}-\delta_{n}}{\bigcirc}
\]

The flip Dynkin Superdiagram

\[
\underset{e_{1}-\delta_{1}}{\otimes}-\underset{\delta_{1}-e_{2}}{\otimes}-\cdots-\underset{e_{m}-e_{m+1}}{\otimes}-\underset{e_{m+1}-\delta_{1}}{\bigcirc}-\underset{\delta_{1}-\delta_{2}}{\otimes}-\cdots-\underset{\delta_{n-1}-\delta_{n}}{\bigcirc}\Rightarrow\underset{\delta_{n-1}-\delta_{n}}{\bigcirc}
\]

case $osp(2,6)$

\[
\underset{e_{1}-e_{2}}{\bigcirc}-\underset{e_{2}-e_{3}}{\bigcirc}-\underset{e_{3}-\delta_{1}}{\otimes}-\underset{\delta_{1}-\delta_{2}}{\bigcirc}-\underset{\delta_{1}-\delta_{2}}{\bigcirc}\Rightarrow\underset{\delta_{2}}{\bigcirc}
\]

\[
\underset{e_{1}-\delta_{1}}{\otimes}-\underset{\delta_{1}-e_{2}}{\otimes}-\underset{e_{2}-\delta_{2}}{\bigcirc}-\underset{\delta_{2}-e{}_{3}}{\otimes}-\underset{e{}_{3}-e_{4}}{\bigcirc}\Rightarrow\underset{e_{4}}{\bigcirc}
\]

case $D(2,1;\alpha$) 

\[
\begin{array}{ccccc}
 &  &  & \bigcirc & 2e_{2}\\
 &  & \diagup\\
e_{1}-e_{2}-e_{3} & \otimes\\
 &  & \diagdown\\
 &  &  & \bigcirc & 2e_{3}
\end{array}
\]

With the normalisation $(e_{1},e_{1})=\frac{-(1+\alpha)}{2}$, $(e_{2},e_{2})=\frac{1}{2}$,
$(e_{3},e_{3})=\frac{\alpha}{2}$ , flip Dynkin Superdiagram becomes
\[
\begin{array}{ccccc}
 &  &  & \otimes & e_{1}+e_{2}-e_{3}\\
 &  & \diagup\\
2e_{2} & \bigcirc\\
 &  & \diagdown\\
 &  &  & \otimes & e_{1}-e_{2}-e_{3}
\end{array}
\]

since $(e_{1}+e_{2}-e_{3},e_{1}+e-e_{3})=0$ and $(e_{1}-e_{2}-e_{3},e_{1}-e-e_{3})=0$.

case $G(3)$  

There is a trivial flip in the dynkin diagram of $G(3)$ to preserve
the orientation and the connected lines between roots. Thus

$\xymatrix{\\
\otimes-\bigcirc\Lleftarrow\bigcirc\ar[r]^{f} & \otimes-\bigcirc\Lleftarrow\bigcirc
}
$

\end{document}